# Superconductivity Induced by High Pressure in Weyl Semimetal TaP

Yufeng Li[1*], Yonghui Zhou[2*], Zhaopeng Guo[1*], Xuliang Chen[2], Pengchao Lu[1], Xuefei Wang[2], Chao An[2], Ying Zhou[2], Jie Xing[1], Guan Du[1], Xiyu Zhu[1,3], Huan Yang[1,3], Jian Sun[1,3†], Zhaorong Yang[2,3†], Yuheng Zhang[2,3] and Hai-Hu Wen[1,3†]

[1] National Laboratory of Solid State Microstructures and Department of Physics, Nanjing University, Nanjing 210093, China

[2] High Magnetic Field Laboratory, Chinese Academy of Sciences, Hefei 230031, China

[3] Collaborative Innovation Center of Advanced Microstructures, Nanjing University, Nanjing, 210093, China

**Weyl semimetal defines a material with three dimensional Dirac cones which appear in pair due to the breaking of spatial inversion or time reversal symmetry. Superconductivity is the state of quantum condensation of paired electrons. Turning a Weyl semimetal into superconducting state is very important in having some unprecedented discoveries. In this work, by doing resistive measurements on a recently recognized Weyl semimetal TaP under pressure up to about 100 GPa, we observe superconductivity at about 70 GPa. The superconductivity retains when the pressure is released. The systematic evolutions of**

**resistivity and magnetoresistance with pressure are well interpreted by the relative shift between the chemical potential and paired Weyl points. Calculations based on the density functional theory also illustrate the structure transition at about 70GPa, the phase at higher pressure may host superconductivity. Our discovery of superconductivity in TaP by pressure will stimulate further study on superconductivity in Weyl semimetals.**

**Introduction**

Superconductivity and topological quantum state are two important fields of frontier research in nowadays condensed matter physics[1-3]. The Weyl semimetal which contains the long sought exotic excitations, namely the Weyl Fermions has been proposed recently[4]. By definition, a Weyl sememetal is characterized by Weyl nodes that possess three dimensional linear band crossing and appear in pair due to the non-centrosymmetric structure or time reversal symmetry breaking. The relevant electronic bands may cross or pass through nearby the so-called Weyl nodes with different chirality and the low-energy excitations near these points disperse linearly in the momentum space. When the Weyl semimetal falls into superconducting state, it is possible to observe some exotic excitations, such as the Majorana fermions. Recently, TaAs and TaP were predicted to be Weyl semimetals[5-6], which is followed immediately by experimental proofs[7-9]. The Weyl semimetal normally exhibits a giant and sometime negative magnetoresistance due to the semimetal behavior and the chirality of the electrons of the paired Weyl nodes[10-11]. Due to the spin chirality of the electrons near the paired Weyl points, it is very interesting and highly desired to realize bulk superconductivity in such Weyl semimetals. Previously, people observed superconducting-like anomaly in the I-V characteristics and regarded it as the observation of superconductivity by pressing a metallic tip on TaAs[12]. In this work, for the first time, we report the discovery of bulk superconductivity in Weyl semimetal TaP by applying high pressure.

**Results**

**Pressure induced evolution of resistivity and magnetoresistance.**

Fig. 1 shows the temperature dependent resistance R(T) of TaP single crystal under different pressures up to 92GPa. Fig. 1**a-b** display the raw data of resistance under pressures from 0.4GPa to 14.6GPa (low pressure region) and from 19.2GPa to 92GPa (high pressure region), respectively. In the whole temperature range, the resistance globally gets enhanced with applied pressure starting from ambient pressure, and reaches a maximum value at about 19.2GPa. Then the resistance decreases monotonically as the pressure continues to rise. In the low pressure region, there is an anomaly showing a maximum of resistance in the intermediate temperature region. This anomaly becomes smeared after the resistance reaches the maximum value at 19.2GPa, and totally disappears in the high pressure region. Three-dimensional contour plots of resistance with temperature and pressure are shown in Fig. 1**c**. The "ridge" like surface graphically shows the variation of resistance with temperature and pressure. Six typical R(T) curves with pressures of 0.4GPa, 5.0GPa, 8.5GPa, 19.2GPa, 46.3GPa and 80.0GPa are representatively plotted in the contour plots. We will see later that this complex evolution is closely related to the crossing of Fermi energy through the paired Weyl points.

Fig. 2**a** shows the change of resistance at 5K and 300K under different pressures. As we can see both of them show similar behavior under external pressure with a peak at around 19.2GPa. The maximum value of resistance at 19.2GPa may be related to the change of Fermi level according to our band structure calculations. In TaP or TaAs families, the Weyl points can be divided into two types according to their positions in reciprocal space: W1 ($k_z$ = 0) and W2 ($k_z$ ≠ 0), these two types of Weyl points locate at different energies[5-6]. For TaP at ambient pressure, the energy of W2 is about 15meV above Fermi level, which gives a finite density of states (DOS) of electrons near the Fermi level. The energy of W2 relative to Fermi level changes significantly with pressure by our calculations. Energies of W2 as a function of pressure from 0GPa to 30GPa and the relative position of Fermi energy are added as cartons in Fig. 3**e**. W2 is above Fermi level at 0GPa and 10GPa, and becomes below the Fermi level when a pressure at around 20GPa is passed. As we can see, when the pressure is increased from the ambient one to about 20GPa, the finite DOS

corresponding to this Weyl point is getting smaller, leading to the increase of the global resistivity. When the pressure is larger than about 20GPa, the reversal process dominates. The W2 point becomes below the Fermi energy and the DOS becomes larger when the pressure is further increased. This leads to the dropping down of the global resistance and a maximum of resistivity in the intermediate temperature region is absent. We calculated electronic resistance of the I4$_1$md structure and the DOS near Fermi level using DFT and the semi-classic Boltzmann transport theory, as shown in Fig. 2**d**, which exhibits the same trend as the experimental data. This picture also gets support from the behavior of magnetoresistance. Fig. 2**b** and 2**c** show the magnetoresistance (MR) at 4.5K and different pressures. A giant magnetoresistance shows up at a low pressure, which was considered as a result of the contribution of the electron and hole pockets in the material[9-10]. The MR effect decreases with applied pressure when this situation is changed and this is very similar to the previous results of giant MR in the semimetal material WTe$_2$[12]. More in detail, from the semi-classical two-band model, the magnetoresistance can be calculated as follows

$$MR = \frac{\Delta\rho}{\rho_0} = \frac{\rho(B) - \rho_0}{\rho_0} = \frac{\sigma_e \sigma_h (\mu_h - \mu_e)^2 B^2}{(\sigma_h + \sigma_e)^2 + (\mu_h \sigma_e + \mu_e \sigma_h)^2 B^2} . \qquad (1)$$

Where $\rho_0$ is the resistivity at zero field, $\sigma_i = n_i e^2 \tau_i / m_i$ (*i=h,e*) is the conductivity (being always positive) at the hole and electron derived band, respectively. Here $\mu_i = e_i \tau_i / m_i$ (*i*=h,e) is the mobility of each band with a negative (positive) sign for the electron (hole) band. At ambient pressure, the Fermi energy is in between W1 and W2 (close to W2), we have an electron and hole contribution for conductivity from the Weyl nodes of W1 and W2 respectively, therefore we have a sizable MR. With compression, the chemical potential gets higher and becomes closer to W2. The electron-pocket Fermi surfaces around W1 become larger while the hole-pocket Fermi surfaces around W2 become smaller. Therefore, it is nature that the MR decreases when increasing pressure. We must emphasize that the MR at 0.4GPa and 8T here is only about 280% which is much smaller than that reported in previous literature[11]. This may be induced by the different directions of the magnetic field and measuring current together with the crystallographic axes. In the pristine sample under zero pressure, we also get about 30000% MR at 3K and 8T (see, Supplementary Fig.2**b**). Note that by applying a high pressure, the giant MR is suppressed and superconductivity is discovered in WTe$_2$ and ZrTe$_5$[13-16]. As follows, we will show that superconducting behavior has also been detected in TaP with further increasing pressure.

**Discovery of superconductivity under pressure.**

When the external pressure reaches 71.0 GPa, the resistance shows a small drop at about 2.3K, as shown in the lower-right inset picture of Fig. 3**a** by an enlarged view in the low temperature region. This drop keeps more pronounced with pressure up to 100GPa. Fig. 3**a** shows the R(T) curves at 92GPa, 96GPa and 100GPa with the pressure increasing procedure, all of them show the resistance drop in low temperature region. In order to check whether this drop corresponds to a superconducting transition, we measure the temperature dependence of resistance under different magnetic fields. The upper inset reveals the resistive transitions at different external magnetic fields up to 1T at 92GPa. It is clear that the transition is gradually suppressed by external magnetic field, this is a typical behavior of a superconductor. Unfortunately, our high-pressure measurement setup cannot be cooled below 1.7K, therefore the superconducting transition cannot be viewed in a complete way. And the maximum pressure value of the experimental setup is about 100GPa, so we cannot further increase the pressure on the sample. Interestingly, when we decrease the pressure this resistance transition becomes much more pronounced. The resistance vs temperature curve measured in the pressure decreasing procedure is shown in Fig. 3**b**. The resistance drop has a maximum with magnitude of about 75% of the normal state value at 30GPa, while it reduces with further lowering pressure, as shown in the left inset of Fig. 3**b**. This experiment rules out any possibility that the superconductivity is arising from the possible impurity phase of Ta after the pressurizing, since Ta is superconductive at about 4.48K at ambient pressure. In addition, because there are no magnetic atoms in TaP so this transition with a drop of about 75% cannot be ascribed to any magnetic transitions. The sharp drop of resistance and the magnetic field dependence demonstrate this transition is a superconducting transition, like that in BaFe$_2$S$_3$[17]. The inset on the right hand side of Fig. 3**b** shows the phase diagram of the onset superconducting transition temperature (T$_c$) with applied pressure. T$_c$ decreases from 2.7K at 30GPa to 1.4K at 0GPa when we reduce the pressure on the sample. The superconductivity retains after pressure released may suggest that TaP transforms to a metastable phase showing superconductivity under a high pressure and it keeps to ambient pressure when the pressure is released.

Since superconductivity survives when the pressure is released, we thus did a new round of pressuring experiment. We chose a new TaP sample and ramp the pressure up to 100GPa, keep 24 hours and then quickly release the pressure to ambient one. Afterwards, we put electrode contacts to measure the resistive transition of this sample at ambient pressure. Since the sample has only a size of about 30 μm after being successfully separated from the diamond anvil cell, we only managed to make two electrodes instead of four on the sample with an irregular shape. But this allows us to measure the superconducting transition down to He$^3$ temperature (T=0.3K). We thus made the resistivity measurement with different magnetic fields down to 0.3K on the sample after the pressure is released to ambient one. Fig. 4**a** shows the resistance under different magnetic fields up to 16T. One can clearly see the superconducting transition starting at about 3.25K. The superconducting transition shifts gradually down to lower temperatures with the magnetic field and the transition is totally suppressed below 0.4K when magnetic field exceeds 4T. A positive and moderate magnetoresistance effect exists here. In addition, we determine the upper critical field $\mu_0H_{c2}$(T) by using the standard crossing method, which determines the superconducting transition temperature by the crossing point of the normal state flat line and the extrapolating line of the steep transition part. In Fig. 4**b** we plot the $\mu_0H_{c2}$ versus *T* by using filled symbols, and the fitting result by the Werthamer-Helfand-Hohenberg (WHH)[18] formula is shown by red dashed line. The experimental data slightly deviates from the WHH formula at low temperature. However, the upper critical field shows a more straight line behavior in the intermediate temperature region, which leads to an extrapolated value of $\mu_0H_{c2}$(0) of about 4-5T.

**Calculation on enthalpy-pressure correlation and structure determination.**

In order to understand the pressure induced evolution of the resistive properties, we have carried out the calculations on the atomic and the associated electronic structures. The technique for the calculation is given in Methods. The resultant enthalpy-pressure and volume-pressure curves are shown in Fig. 5**b**. Interestingly, with compression, a structure phase transition occurs at around 70GPa, from the ambient pressure structure I41md (space group No.109) to a Pmmn (space group No.59) phase. The crystal structures of all the stable TaP phases are shown in Fig. 5**a**, while it is well known that the ambient-pressure phase is a face-centered tetragonal structure and the coordination number of Ta atoms in this phase is six. It is interesting that, there are two different coordination numbers for the Ta atoms in the Pmmn phase, seven for Ta atoms at Wyckoff Position 2b (0,0,0.3326) and five for other two Ta atoms at position 2a (0,0,0.0826). With increasing pressure, crystal structures tend to become more compacted. As shown in Fig. 5**b**, clear volume collapsing is predicted along with the structure phase transitions.

**Discussion**

In order to know the origin of superconductivity, we need to investigate the electronic structures of the Pmmn phase under pressure of 70GPa or above. We have done the first principle calculations, the results are shown in Fig. **6**. For the electron band of the phase at and above 70GPa, there are six bands across the Fermi level in X-Γ direction, which implies the metallicity of the Pmmn phase, this is consistent with the experiment. In order to discuss the Fermi surface nesting, we plot Fermi surface contour at the $k_z$=0 plane at 70GPa, 40GPa and 0GPa which are shown in Fig. **6c-e**, respectively. In the contour plot of Fermi surface at 70GPa, there are four electron pockets around $(k_x, k_y) = (0.3, 0.3) \times 2\pi$, $(k_x, k_y) = (0.3, -0.3) \times 2\pi$, $(k_x, k_y) = (-0.3, 0.3) \times 2\pi$ and $(k_x, k_y) = (-0.3, -0.3) \times 2\pi$ connected with mirror and rotation symmetry. The electron-like Fermi surface around $(k_x, k_y) = (0.3, 0.3) \times 2\pi$ will nest with the Fermi surface around $(k_x, k_y) =$

$(-0.3,-0.3) \times 2\pi$ with the q-vector $\vec{q} = (0.6, 0.6) \times 2\pi$. The contour plots of Fermi surface in 40GPa and 0GPa after releasing from a high pressure have similar feature as that of 70GPa.

The existence of Fermi surface nesting suggests possible superconducting properties in this phase. According to the Bardeen-Copper-Schrieffer theory, the pairing gap is described by $\Delta = 2\hbar\omega_D \exp(-1/N(0)V)$, here $\omega_D$ is the Debye frequency, *N(0)* is the density states at Fermi energy, and *V* is the attractive interaction through exchanging a phonon. When we have a moderate density of states with the suitable electron-phonon coupling, superconductivity may be achieved. The calculations suggest that when the pressure is released from this high-pressure Pmmn phase, the superconductivity may still appear due to the quite similar electronic structures before and after releasing pressure. This gives a reasonable explanation to our data. At this moment, we are not able to see the structure predicted by our calculation under high pressure since it is really challenging to do structure measurement when the pressure is above 70GPa. In addition, it is highly desired to know how much feature of the Weyl semimetal is still carried by the phase after pressuring under pressures above 70GPa. Therefore superconductivity discovered directly from pressurized Weyl semimetal TaP will stimulate further efforts in investigating superconductivity in topological materials and in helping the exploration of the Majorana Fermions.

# Methods

**I. Sample growth**

TaP single crystals were synthesized by chemical vapor transport (CVT) method which is similar to the previous work[19]. About 4g polycrystalline TaP precursors were prepared first by sintering tantalum and phosphorus powders in vacuum in quartz ampoules at 973K. Then the precursors with 15mg/cm$^3$ iodine were sealed in a 30cm long quartz ampoule under vacuum. The ampoule was heated to a temperature with a gradient between 950°C and 850°C, with the starting materials at the high temperature side. Single crystals were obtained by transferring after staying with this temperature gradient for three weeks.

**II. High Pressure measurements**

High-pressure resistance measurements with different cycles were conducted via the four probe method in a screw-pressure-type diamond anvil cell (DAC). Platinum (Pt) foil with a thickness of 2μm was used as electrode. Diamond anvils of 200-μm culets and rhenium gasket covered with a mixture of epoxy and fine cubic boron nitride (c-BN) powder were used here. A single crystal with typical dimension of 65×20×5 μm$^3$ was loaded without pressure-transmitting medium. Pressure was calibrated by using the ruby fluorescence scale below 70 GPa[20] and the diamond Raman scale above 70 GPa[21].

## III. Theoretical Calculations

Ab initio random searching (AIRSS) method[22,23] was applied for crystal structure predictions in the frame of density-functional theory (DFT). Enthalpy and band structures calculations were performed with the projected augmented wave (PAW) method implemented in the Vienna ab initio simulation package code (VASP)[24-26] with Perdew-Burke-Ernzerhof (PBE)[27] generalized gradient approximation (GGA) exchange-correlation functional. The cutoff energy of 400 eV was set for a plane wave basis. The calculations of phonon spectrum were performed in a $3 \times 3 \times 2$ supercell being computed using the PHONOPY code[28] together with VASP code.

The calculations of semi-classic transport conductivity were performed using Boltzmann equation method implemented in the BoltzTraP package[29-30], and the mesh for self-consistent band energies is 16x16x22. Spin-orbit coupling was concerned during calculations.

## Acknowledgements

This work was supported by the Ministry of Science and Technology of China (Grant No. 2016YFA0300400, 2015CB921202, 2016YFA0401804), the national natural science foundation of China (NSFC) with the projects: 11534005, 11190023, 11374143, U1532267, 51372112, 11574133, 11574323, U1632275, NSF Jiangsu province (No. BK20150012), the Special Program for Applied Research on Super Computation of the NSFC-Guangdong Joint Fund

(the second phase), and HPCC of Nanjing University.

**Reference**

1. Qi, X. L. & Zhang, S. C. Topological insulators and superconductors. *Rev. Mod. Phys.* **83**, 1057 (2011).

2. Nadj-Perge, S. *et al.* Observation of Majorana fermions in ferromagnetic atomic chains on a superconductor. *Science* **346**, 602-607 (2014).

3. Masatoshi Sato, and Yoichi Ando, Topological Superconductors. arXiv.1608.063395.

4. Wan, X. G. et al. Topological semimetal and Fermi-arc surface states in the electronic structure of pyrochlore iridates. *Phys. Rev. B* **83**, 205101(2011).

5. Weng, H. M. *et al.* Weyl semimetal phase in non-centrosymmetric transition metal monophosphides. *Phys. Rev. X* **5**, 011029 (2015).

6. Huang, S. M. *et al.* Theoretical Discovery/Prediction: Weyl Semimetal states in the TaAs material (TaAs, NbAs, NbP, TaP) class, *Nature Commun.* **6**, 7373 (2015)

7. Xu, S. -Y. *et al.* Experimental discovery of a topological Weyl semimetal state in TaP. *Science Advances* **1**, e1501092 (2015).

8. Xu, N. *et al.* Observation of Weyl nodes and Fermi arcs in TaP. *Nature Commun* **7**, 11006 (2016).

9. Inoue, H et al. Quasiparticle interference of the Fermi arcs and surface-bulk connectivity of a Weyl semimetal. *Science* **351**, 1184 (2016)

10. Arnold, F. *et al.* Negative magnetoresistance without well-defined chirality in the Weyl semimetal TaP. *Nature Commun.* **7**, 11615 (2016).

11. Du, J. *et al.* Unsaturated both large positive and negative magnetoresistance in Weyl Semimetal TaP. *Sci. China-Phys. Mech. Astron.* 59, 657406(2016).

12. Wang, H. *et al.* Tip induced unconventional superconductivity on Weyl semimetal TaAs. arXiv:1607.00513.

13. Ali, M. N. *et al.* Large, non-saturating magnetoresistance in $WTe_2$. *Nature* **514**, 205–208 (2014).

14. Pan, X. C. *et al.* Pressure-driven dome-shaped superconductivity and electronic structural evolution in tungsten ditelluride. *Nature Commun.* **6**, (2015).

15. Kang, D. F. *et al.* Superconductivity emerging from a suppressed large magnetoresistant state in tungsten ditelluride. *Nature Commun.* **6** (2015).

16. Zhou, Y. H. *et al.* Pressure induced superconductivity in a three dimensional topological material $ZrTe_5$. *PNAS* 113, 2904 (2016).

17. Takahashi, H. et al. Pressure-induced superconductivity in the iron-based ladder material $BaFe_2S_3$. *Nature materials* **14**, 1008 (2015).

18. Werthamer, N. R. *et al.* Temperature and purity dependence of the superconducting critical field $H_{c2}$. III. Electron spin and spin-orbit effects.

*Phys. Rev.* **147**, 295 (1966).

19. Du, J. *et al.* Large unsaturated positive and negative magnetoresistance in Weyl semimetal TaP. *Sci. China Phys. Mech. Astron.* **59**, 657406(2016).

20. Mao, H. K. *et al.* Calibration of the ruby pressure gauge to 800 kbar under quasi‐hydrostatic conditions. *J. Geophys. Res.* **91**, 4673 (1986).

21. Akahama, Y. and Kawamura, H. High-pressure Raman spectroscopy of diamond anvils to 250 GPa: Method for pressure determination in the multimegabar pressure range. *J. Appl. Phys.* **96**, 3748 (2004).

22. Pickard, C. J. *et al.* High-Pressure Phases of Silane. *Phys. Rev. Lett.* **97**, 045504 (2006).

23. Pickard, C. J. *et al.* Ab initio random structure searching. *J. Phys.: Condens. Matter* **23**, 053201 (2011).

24. Kresse, G. *et al.* From ultrasoft pseudopotentials to the projector augmented-wave method. *Phys. Rev. B* **59**, 1758 (1999).

25. Kresse, G. *et al.* Efficient iterative schemes for ab initio total-energy calculations using a plane-wave basis set. *Phys. Rev. B* **54**, 11169 (1996).

26. Kresse, G. *et al.* Efficiency of ab-initio total energy calculations for metals and semiconductors using a plane-wave basis set. *Comput. Mater. Sci.* **6**, 15-50 (1996).

27. Perdew, J. P. *et al.* Generalized Gradient Approximation made Simple. *Phys. Rev. Lett.* **77**, 3865 (1996).

28. Togo, A. *et al.* First-principles calculations of the ferroelastic transition

between rutile-type and CaCl2-type SiO2 at high pressures. *Phys. Rev. B* **78**, 134106 (2008).

29. Madsen, G. K. *et al.* A code for calculating band-structure dependent quantities, *Comput. Phys. Commun.* **175**, 67 (2006).

30. Kuneš J. *et al.* Electronic structure of fcc Th: Spin-orbit calculation with 6p 1/2 local orbital extension. *Phys. Rev. B* **64**, 153102 (2001).

## Author Contributions

The samples were grown by Y.F.L and X.Y.Z. The high pressure resistivity measurements were conducted by Y.H.Z., X.L.C., X.F.W., C.A., Y.Z. and Z.R.Y. The low temperature with He$^3$ was measured by J.X., Y.F.L., G.D. and H.Y. The DFT calculations were done by Z.P.G., P.C.L., and J.S. H-H.W. coordinated the whole work. H.-H.W, Y.F.L and J.S. wrote the manuscript with the supplementary by others. All authors have discussed the results and the interpretations.

## Author Information

The authors declare no competing financial interests. Correspondence and requests for materials should be addressed to H-H.W, Z.R.Y. and J.S. (hhwen@nju.edu.cn, zryang@issp.ac.cn, jiansun@nju.edu.cn).

**Figures and legends**

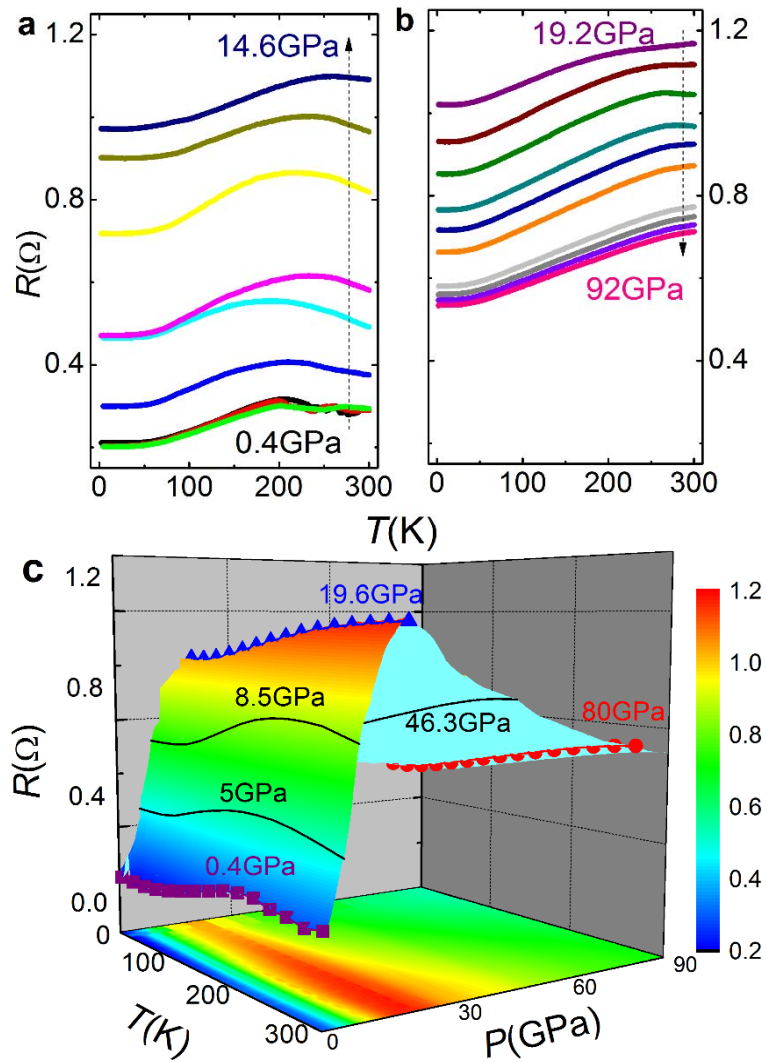

**Figure 1 | Temperature-dependent resistance under pressures.** Shown in a is the temperature dependence of resistance for pressures from 0.4GPa to 14.6GPa, and those for pressures from 19.2GPa to 92.0GPa are shown in b. The three-dimensional contour plots of resistance versus temperature under different pressures is shown in c. The R(T) curves at six typical pressures are also plotted in c to indicate the change of resistance with temperature and pressure.

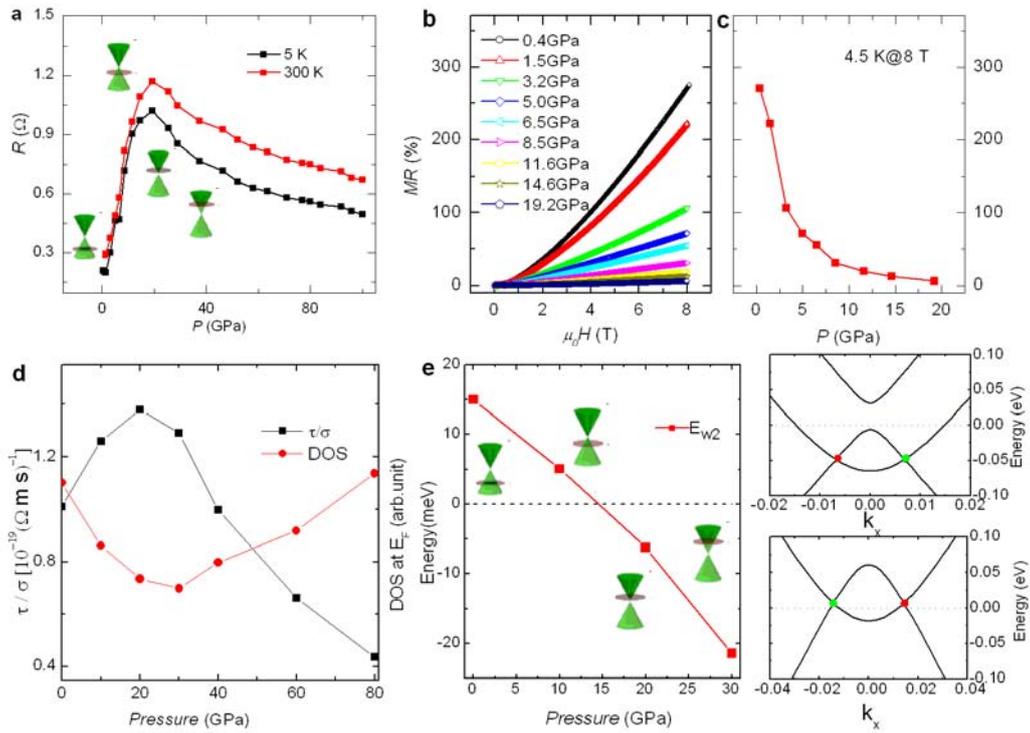

**Figure 2 | Resistance and magnetoresistance under pressure.** Resistance at 5K and 300K versus pressure are shown in **a**. Both of them show a maximum value at about 19.2GPa. Four insets in **a** indicate the change of Fermi level relative to W2 from 0GPa to 30GPa through band structure calculations. The Fermi level crosses the Weyl points between 10GPa and 20GPa, which corresponds to the maximum value of resistance at pressures around 20GPa. Shown in **b** and **c** are the MR versus magnetic field and pressure, respectively. In **d** we show the calculated semi-classic transport resistance (black circles) and the DOS near Fermi level (red circles) of the $I4_1md$ phase vs. pressure. **e** Energies of W2 relative to the Fermi energy under different pressures and local band structures of two kinds of Weyl points in the $I4_1md$ phase. The Dirac cones are schematically shown together with the Fermi level (grey disc) under different pressures. The red line shows the energy of W2 evolution along with pressure. Weyl points are denoted by red and green dots to represent opposite chirality.

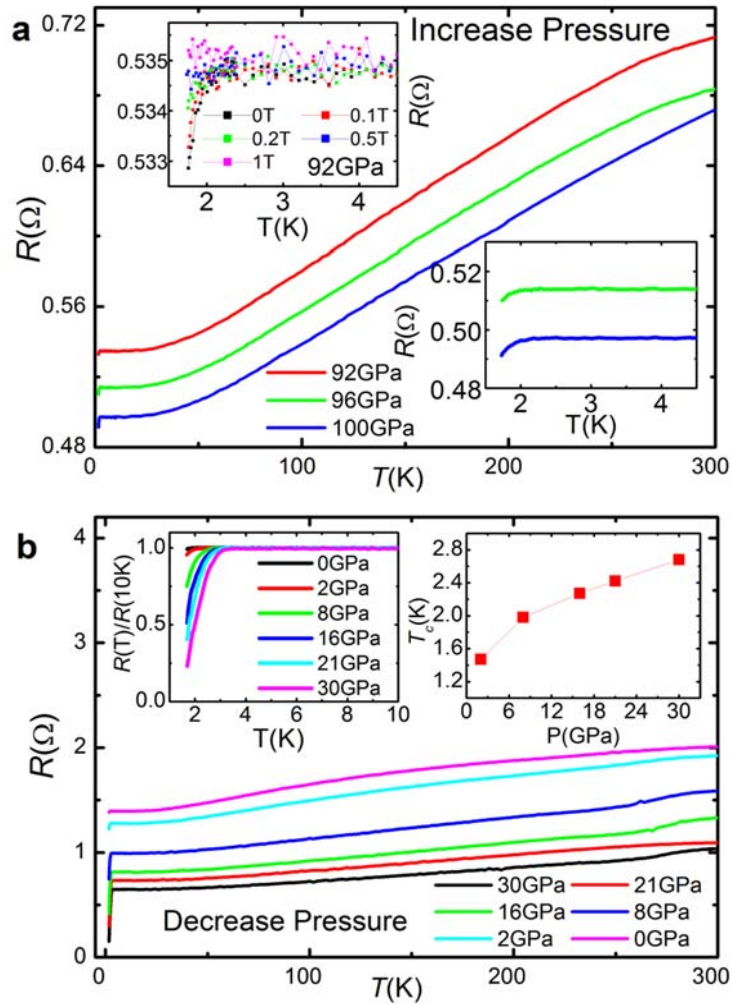

**Figure 3 | Temperature dependent resistance with increasing and decreasing pressures.** Shown in **a** are the R(T) curves from 92GPa to 100GPa with increasing pressure. The R(T) curves under different pressures with decreasing pressure are shown in **b**. With increasing pressure there exists a sudden drop of resistance and this becomes clearer after releasing pressure. This transition is ascribed to the superconducting transition induced by pressure, and the superconductivity retains when the pressure is released to ambient one. The upper-left inset in **a** shows the resistance versus temperature under different fields and the lower-right inset shows the enlarged view of the data measured under 71GPa in low temperature region. About 75% drop of the normal state resistance under 30GPa can be seen in the left inset in **b.** The inset on the right hand side shows the pressure dependence of $T_c$ in decreasing

pressure.

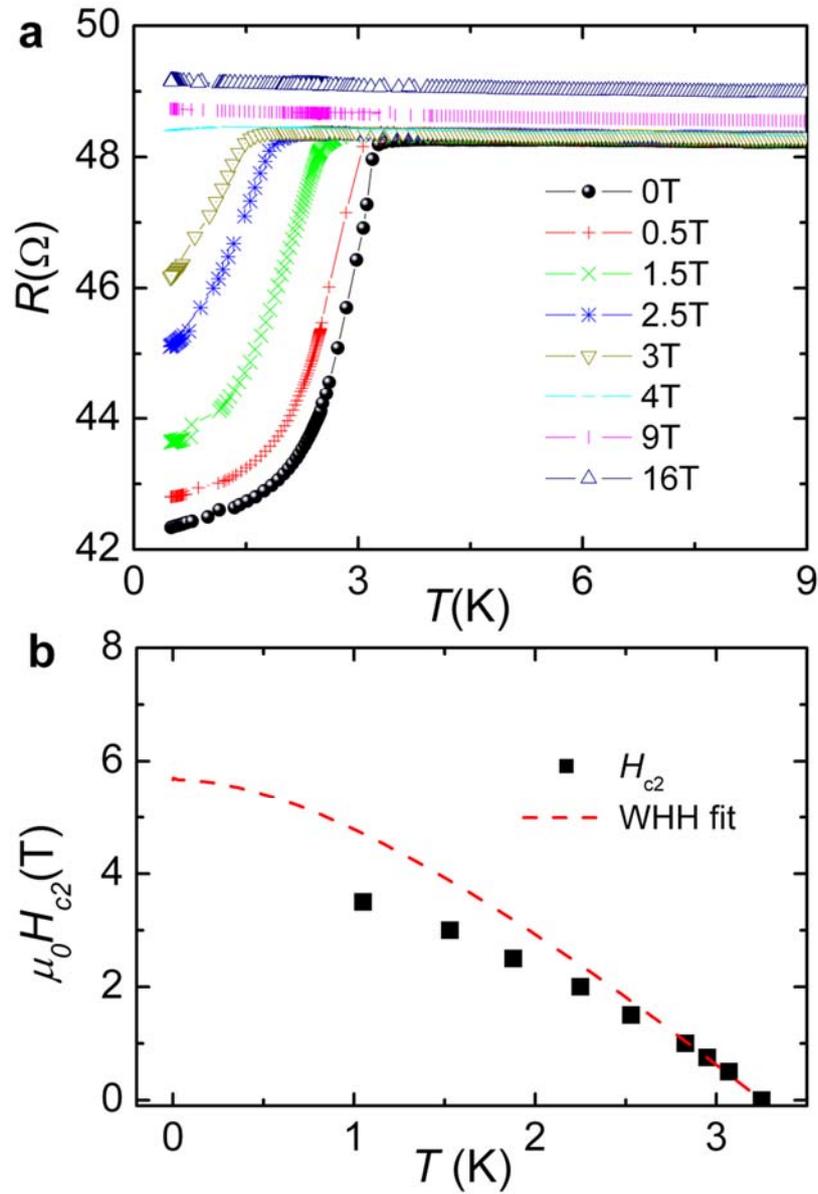

**Figure 4 | Temperature dependence of resistance and upper critical field when the pressure is released to ambient. a,** R(T) curves under fields up to 16T. The superconducting transition is totally suppressed above 0.4K at 4T. **b,** The upper critical field $\mu_0H_{c2}$ versus $T$ shown by the black squares. The red dashed line is the WHH fitting curve.

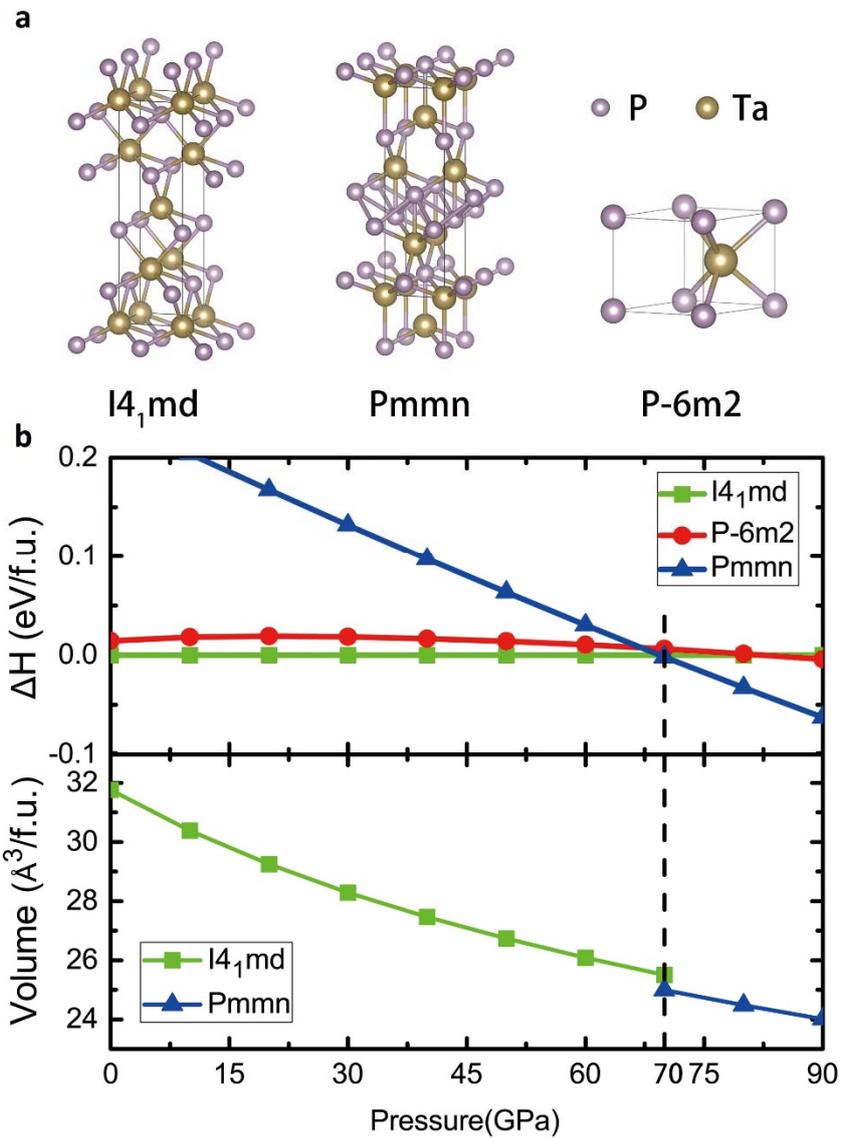

**Figure 5 | Predicted phases and structures of TaP and calculated enthalpy and volume.** (a) Crystal structures of I4$_1$md, Pmmn and P-6m2 phases. The balls in violet and golden represent P and Ta atoms, respectively. (b) Enthalpies per formula of TaP phases relative to the I4$_1$md phase vs pressure, and volume-pressure curves for I4$_1$md and Pmmn structures. Both enthalpy and volume are normalized to that of per formula of TaP.

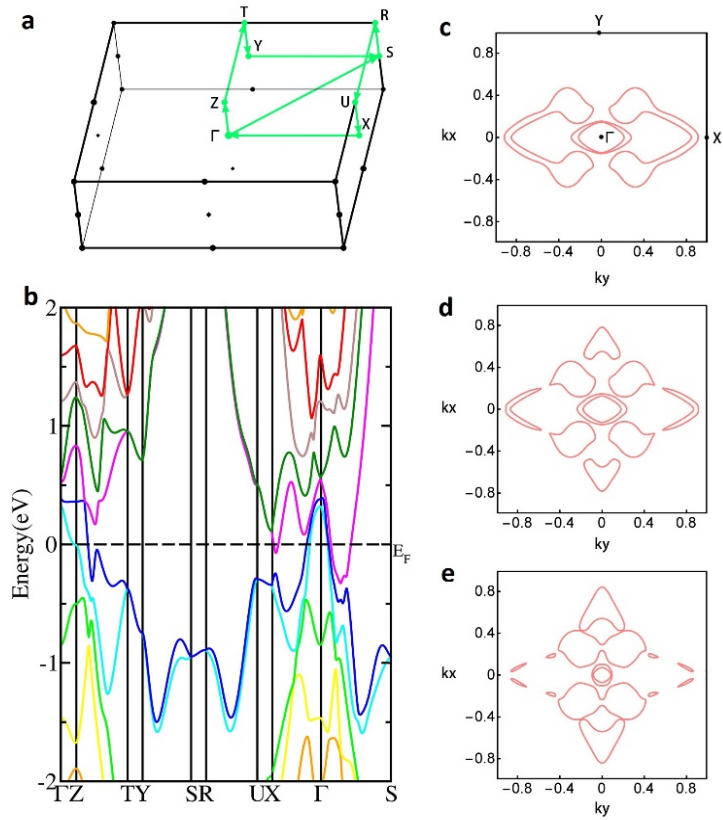

**Figure 6 | Brillouin zone and electronic structures of the TaP Pmmn phase.**
(a) The Frist Brillouin zone and the k-points' path of TaP Pmmn phase. (b) Band structures of the Pmmn phase with SOC at 70GPa. And contour plots of Fermi surfaces at 70GPa (c), 40GPa (d) and 0GPa (e) for the Pmmm phase.

# Supplementary Information

# Superconductivity Induced by High Pressure in Weyl Semimetal TaP

Yufeng Li[1*], Yonghui Zhou[2*], Zhaopeng Guo[1*], Xuliang Chen[2], Pengchao Lu[1], Xuefei Wang[2], Chao An[2], Ying Zhou[2], Jie Xing[1], Guan Du[1], Xiyu Zhu[1,3], Huan Yang[1,3], Jian Sun[1,3†], Zhaorong Yang[2,3†], Yuheng Zhang[2,3] and Hai-Hu Wen[1,3†]

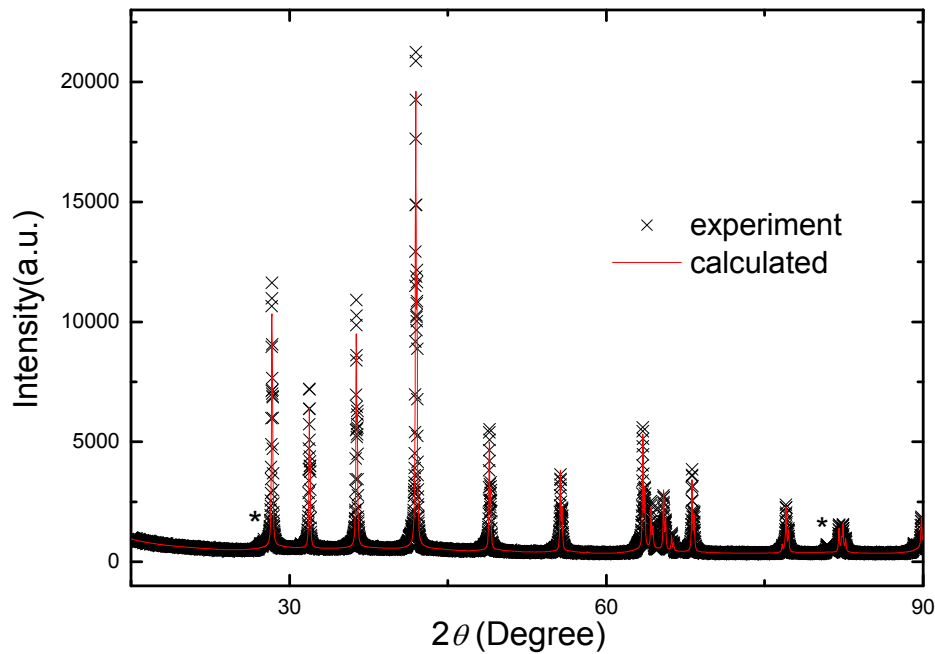

**Supplementary Figure 1 | The X-ray diffraction pattern of TaP powders.** The red line is the Rietveld refinement result of TaP at room temperature. Two peaks marked with asterisk (*) may come from the instrumental background.

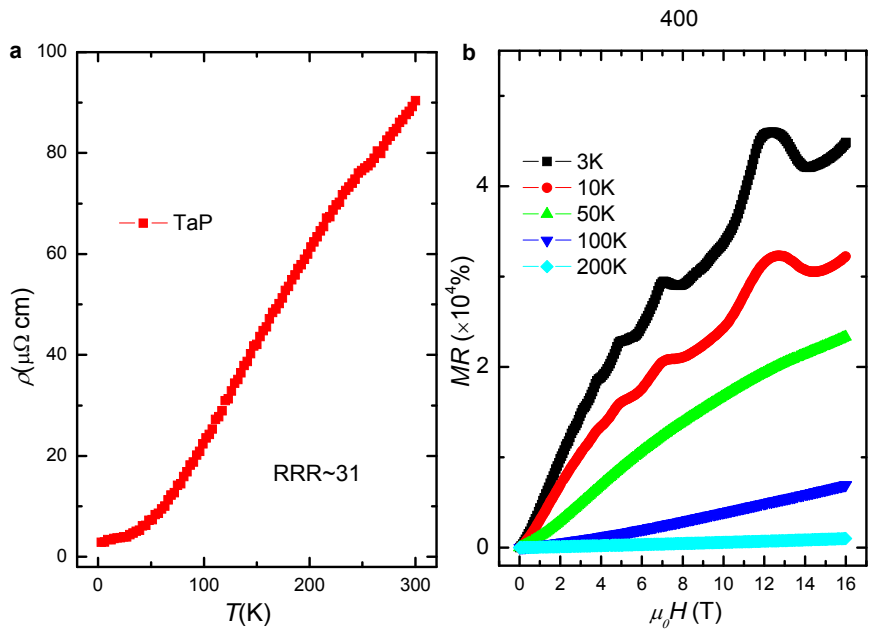

**Supplementary Figure 2 | Resistivity and magnetoresistance (MR) of TaP single crystals**. (a) Temperature dependent resistivity under ambient pressure of TaP. (b) MR data at different temperatures with external magnetic field paralleled to c-axis. The SdH oscillations appear on the MR curve at 3K (black) and 10K (red).

I. XRD characterization and magnetoresistance of the pristine TaP sample

The XRD data were collected on TaP powders by crushing and grinding the as prepared TaP single crystals. The XRD and the Rietveld refinement result with weighted profile R-factor $R_{wp}$ = 8.789% are shown in Supplementary Figure 1. The refined cell parameters are a=3.320Å and c=11.343Å. Two tiny peaks that cannot be indexed with TaP may come from the instrument background. The rather clean XRD data indicate the high purity of the samples.

Temperature dependent resistivity of a TaP single crystal under ambient pressure is shown in the left panel of Supplementary Fig. 2. The residual resistivity ratio (RRR) is about 31. Magnetoresistance (MR) of the pristine TaP single crystal is shown in the right panel of Supplementary Fig. 2. The MR data were taken at 3K, 10K, 50K, 100K and 200K up to 16T. Shubnikov-deHaas (SdH) oscillations are clearly shown at 3K and 10K. MR gets its maximum value of about 45000% at 3K and 16T.

## II. The evolution of the structure and the Weyl point positions with the pressure

For the ambient-pressure I4$_1$md structure, the energy of W1 is 50meV under the Fermi level and that of W2 is about 15meV above Fermi level, showed in Fig. 2e. We focus on the energy of W2 at different pressures, which may have larger influence on the density of states near the Fermi level and thus the electrical conductibility than W1.

Our calculated coordinates and energies of W2 under pressure from 0GPa to 30GPa are shown in Table 1. Other W2 points are related to the W2 point in Table 1 by mirror symmetry, C$_4$ rotation and time-reversal symmetry. And all the energies of W2 points are equal. The coordinates of W2 at 0GPa are consistent with previous work[S1]. With the increase of pressure, the coordinate x and y of

W2 are basically maintained while the coordinate z of W2 decreases from 0.585 to 0.541. It is surprising that the energies of W2 relative to Fermi level change significantly with the increase of pressure.

Energies of W2 as a function of pressure and relative position of Fermi energy and energies of W2 are shown in Fig. 2e, the green cones are Weyl points with linear dispersion relation and the brown circular planes represent Fermi energy. As the red line shows, with the increase of pressure, the energy of W2 decrease almost linearly, and across the Fermi energy at around 15GPa. The density of states will be local minimum related to pressure around 15GPa, and derives a local minimum of electrical conductibility, which agrees the experimental measurements.

Table 1 | The coordinates (in units of the length of Γ−Σ for x and y and of the length of Γ-Z for z) and energies (in meV) of W2 in pressure of 0GPa to 30GPa. The energy values are relative to the Fermi level.

|  | x | y | z | E-Weyl2(meV) |
|---|---|---|---|---|
| 0GPa | 0.031 | 0.505 | 0.583 | 15.0 |
| 10GPa | 0.031 | 0.504 | 0.568 | 5.1 |
| 20GPa | 0.030 | 0.503 | 0.553 | -0.62 |
| 30GPa | 0.030 | 0.501 | 0.541 | -21.4 |

**Reference**

S1. Weng, H. *et al.* Weyl semimetal phase in non-centrosymmetric transition metal monophosphides. Phys. Rev. **X** 5, 011029 (2015).